\input harvmac

%% redefine listrefs so that it doesn't  force page break
\def\listrefs{\vbox{}\goodbreak\bigskip\immediate\closeout\rfile\writestoppt
\baselineskip=\footskip\centerline{{\bf References}}\bigskip{\parindent=20pt%
\frenchspacing\escapechar=` \input \jobname.refs\vfill\eject}\nonfrenchspacing}

%%%%%%%%%%%% Macros %%%%%%%%%%%%

\font\tenmsb=msbm10
\font\sevenmsb=msbm7
\font\fivemsb=msbm5
\def\bb{
\textfont1=\tenmsb
\scriptfont1=\sevenmsb
\scriptscriptfont1=\fivemsb
}

\def\C{\hbox{$\bb C$}}
\def\H{\hbox{$\bb H$}}
\def\Z{\hbox{$\bb Z$}}

\def\cN{{\cal N}}
\def\cP{{\cal P}}

\def\SU{\hbox{SU}}
\def\SO{\hbox{SO}}
\def\U{\hbox{U}}
\def\new{^{new}}
\def\hkq{\,/\!/\!/\,}
\def\Higgs{\hbox{Higgs}}

%%%%%%%%%%%% References %%%%%%%%%%%%

%\AharonyDJ
\lref\AharonyDJ{
  O.~Aharony and Y.~Tachikawa,
  ``A holographic computation of the central charges of $d=4$, $\cN=2$ SCFTs,''
  JHEP {\bf 0801}, 037 (2008)
  [arXiv:0711.4532 [hep-th]].
  %%CITATION = JHEPA,0801,037;%%
}

%\ArgyresEH
\lref\ArgyresEH{
  P.~C.~Argyres, M.~R.~Plesser and N.~Seiberg,
  ``The Moduli Space of $\cN=2$ SUSY {QCD} and Duality in $\cN=1$ SUSY {QCD},''
  Nucl.\ Phys.\  B {\bf 471}, 159 (1996)
  [arXiv:hep-th/9603042].
  %%CITATION = NUPHA,B471,159;%%
}

%\ArgyresJJ
\lref\ArgyresJJ{
  P.~C.~Argyres and M.~R.~Douglas,
  ``New phenomena in SU(3) supersymmetric gauge theory,''
  Nucl.\ Phys.\  B {\bf 448}, 93 (1995)
  [arXiv:hep-th/9505062].
  %%CITATION = NUPHA,B448,93;%%
}

%\ArgyresCN
\lref\ArgyresCN{
  P.~C.~Argyres and N.~Seiberg,
  ``S-duality in $\cN=2$ supersymmetric gauge theories,''
  JHEP {\bf 0712}, 088 (2007)
  [arXiv:0711.0054 [hep-th]].
  %%CITATION = JHEPA,0712,088;%%
}

%\ArgyresXN
\lref\ArgyresXN{
  P.~C.~Argyres, M.~Ronen Plesser, N.~Seiberg and E.~Witten,
  ``New $\cN=2$ Superconformal Field Theories in Four Dimensions,''
  Nucl.\ Phys.\  B {\bf 461}, 71 (1996)
  [arXiv:hep-th/9511154].
  %%CITATION = NUPHA,B461,71;%%
}

%\ArgyresFW
\lref\ArgyresFW{
  P.~C.~Argyres and A.~D.~Shapere,
  ``The Vacuum Structure of $\cN=2$ SuperQCD with Classical Gauge Groups,''
  Nucl.\ Phys.\  B {\bf 461}, 437 (1996)
  [arXiv:hep-th/9509175].
  %%CITATION = NUPHA,B461,437;%%
}

%\BarnesJJ
\lref\BarnesJJ{
  E.~Barnes, K.~A.~Intriligator, B.~Wecht and J.~Wright,
  ``Evidence for the strongest version of the 4d $a$-theorem, via $a$-maximization
  along RG flows,''
  Nucl.\ Phys.\  B {\bf 702}, 131 (2004)
  [arXiv:hep-th/0408156].
  %%CITATION = NUPHA,B702,131;%%
}

%\CardyCWA
\lref\CardyCWA{
  J.~L.~Cardy,
  ``Is there a $c$-theorem in four-dimensions?,''
  Phys.\ Lett.\  B {\bf 215}, 749 (1988).
  %%CITATION = PHLTA,B215,749;%%
}

%\ChacaltanaKS
\lref\ChacaltanaKS{
  O.~Chacaltana and J.~Distler,
  ``Tinkertoys for Gaiotto Duality,''
  arXiv:1008.5203 [hep-th].
  %%CITATION = ARXIV:1008.5203;%%
}

%\EguchiVU
\lref\EguchiVU{
  T.~Eguchi, K.~Hori, K.~Ito and S.~K.~Yang,
  ``Study of $\cN=2$ Superconformal Field Theories in $4$ Dimensions,''
  Nucl.\ Phys.\  B {\bf 471}, 430 (1996)
  [arXiv:hep-th/9603002].
  %%CITATION = NUPHA,B471,430;%%
}

%\GaiottoNZ
\lref\GaiottoNZ{
  D.~Gaiotto, A.~Neitzke and Y.~Tachikawa,
  ``Argyres-Seiberg duality and the Higgs branch,''
  Commun.\ Math.\ Phys.\  {\bf 294}, 389 (2010)
  [arXiv:0810.4541 [hep-th]].
  %%CITATION = CMPHA,294,389;%%
}

%\GaiottoWE
\lref\GaiottoWE{
  D.~Gaiotto,
  ``$\cN=2$ dualities,''
  arXiv:0904.2715 [hep-th].
  %%CITATION = ARXIV:0904.2715;%%
}

%\GaiottoHG
\lref\GaiottoHG{
  D.~Gaiotto, G.~W.~Moore and A.~Neitzke,
  ``Wall-crossing, Hitchin Systems, and the WKB Approximation,''
  arXiv:0907.3987 [hep-th].
  %%CITATION = ARXIV:0907.3987;%%
}

%\HananyNA
\lref\HananyNA{
  A.~Hanany and Y.~Oz,
  ``On the Quantum Moduli Space of Vacua of $\cN=2$ Supersymmetric $\SU(N_c)$
  Gauge Theories,''
  Nucl.\ Phys.\  B {\bf 452}, 283 (1995)
  [arXiv:hep-th/9505075].
  %%CITATION = NUPHA,B452,283;%%
}

%\IntriligatorJJ
\lref\IntriligatorJJ{
  K.~A.~Intriligator and B.~Wecht,
  ``The exact superconformal R-symmetry maximizes a,''
  Nucl.\ Phys.\  B {\bf 667}, 183 (2003)
  [arXiv:hep-th/0304128].
  %%CITATION = NUPHA,B667,183;%%
}

%\MinahanFG
\lref\MinahanFG{
  J.~A.~Minahan and D.~Nemeschansky,
  ``An $\cN = 2$ superconformal fixed point with $E_6$ global symmetry,''
  Nucl.\ Phys.\  B {\bf 482}, 142 (1996)
  [arXiv:hep-th/9608047].
  %%CITATION = NUPHA,B482,142;%%
}

%\MyersXS
\lref\MyersXS{
  R.~C.~Myers and A.~Sinha,
  ``Seeing a $c$-theorem with holography,''
  Phys.\ Rev.\  D {\bf 82}, 046006 (2010)
  [arXiv:1006.1263 [hep-th]].
  %%CITATION = PHRVA,D82,046006;%%
}

%\SeibergRS
\lref\SeibergRS{
  N.~Seiberg and E.~Witten,
  ``Monopole condensation and confinement in $\cN=2$ supersymmetric Yang-Mills
  theory,''
  Nucl.\ Phys.\  B {\bf 426}, 19 (1994)
  [Erratum-ibid.\  B {\bf 430}, 485 (1994)]
  [arXiv:hep-th/9407087].
  %%CITATION = NUPHA,B426,19;%%
}
%\SeibergAJ
\lref\SeibergAJ{
  N.~Seiberg and E.~Witten,
  ``Monopoles, duality and chiral symmetry breaking in $\cN=2$ supersymmetric
  QCD,''
  Nucl.\ Phys.\  B {\bf 431}, 484 (1994)
  [arXiv:hep-th/9408099].
  %%CITATION = NUPHA,B431,484;%%
}

%\ShapereZF
\lref\ShapereZF{
  A.~D.~Shapere and Y.~Tachikawa,
  ``Central charges of $\cN=2$ superconformal field theories in four dimensions,''
  JHEP {\bf 0809}, 109 (2008)
  [arXiv:0804.1957 [hep-th]].
  %%CITATION = JHEPA,0809,109;%%
}
%\ShapereUN
\lref\ShapereUN{
  A.~D.~Shapere and Y.~Tachikawa,
  ``A counterexample to the `$a$-theorem',''
  JHEP {\bf 0812}, 020 (2008)
  [arXiv:0809.3238 [hep-th]].
  %%CITATION = JHEPA,0812,020;%%
}

%\draftmode

\Title{IPMU10-0202}{Comments on scaling limits of 4d $\cN=2$ theories}

\bigskip

\centerline{Davide Gaiotto, Nathan Seiberg,\footnote{$^\bullet$}{dgaiotto, seiberg@ias.edu}}

\medskip

\centerline{School of Natural Sciences,
Institute for Advanced Study,}
\centerline{Einstein Drive, Princeton, NJ 08540, USA}

\bigskip

\centerline{and Yuji Tachikawa\footnote{$^\circ$}{yujitach@ias.edu or yuji.tachikawa@ipmu.jp, on leave from the Institute for Advanced Study}}
\medskip
\centerline{Institute for the Physics and Mathematics of the University,
University of Tokyo,}
\centerline{Kashiwa, Chiba 277-8583, Japan}

\bigskip
\bigskip
\bigskip

\noindent
We revisit the study of the maximally singular point in the Coulomb branch of 4d $\cN=2$ $\SU(N)$ gauge theory with $N_f=2n$  flavors for $N_f<2N$. When $n \ge 2$, we find that the low-energy physics is described by two non-trivial superconformal field theories coupled to a magnetic $\SU(2)$ gauge group which is infrared free.  (In the special case $n=2$, one of these theories is a theory of free  hypermultiplets.) This observation removes a possible counter example to a conjectured $a$-theorem.

\bigskip
\bigskip

\Date{November 2010}

\eject

\newsec{Introduction}

$\cN=2$ supersymmetric gauge theory has been a useful playground where non-perturbative dynamics can be studied exactly, thanks to the holomorphy inherent in its low-energy Lagrangian.
For example, monopoles can be made massless in the strongly-coupled region \refs{\SeibergRS,\SeibergAJ}.

It is also possible that both electric and magnetic particles become massless simultaneously, as  was first realized  by Argyres and Douglas \ArgyresJJ\ in the pure $\SU(3)$ gauge theory at a point on the Coulomb branch.
The system at that point is an isolated superconformal theory; here the adjective `isolated' signifies the fact that the theory has no marginal coupling.
General properties of such $\cN=2$ superconformal theories were studied in \ArgyresXN, which also provided realization of such theories in terms of $\SU(2)$ theories with $N_f=1,2$, or $3$ flavors.

A more conventional way to have $\cN=2$ superconformal symmetry is to start with an $\cN=2$ gauge theory whose one-loop beta function vanishes.
Typical examples are $\SU(n)$ gauge theory with $N_f=2n$ flavors.
The complexified gauge coupling $\tau$ of such a theory is exactly marginal and can be tuned.
The strong-coupling limit of such a theory has a dual description consisting of an isolated superconformal theory with a flavor symmetry whose subgroup is gauged by weakly-coupled vector multiplets \refs{\ArgyresCN,\GaiottoWE}.
One important lesson is that the strongly-coupled theory can consist of several distinct strongly-interacting sectors weakly coupled by a gauge field.

For example, the $\tau\to 1$ limit of $\SU(3)$ theory with six flavors consists \ArgyresCN\ of a weakly-coupled $\SU(2)$ gauge multiplet coupled to two sectors: one is a  hypermultiplet in the doublet representation and another is an isolated superconformal theory with $E_6$ flavor symmetry found by \MinahanFG.
The flavor symmetry $\U(6)$ originally carried by six flavors is, in the dual description, realized as the $\U(1)$ symmetry acting on a doublet hypermultiplet and the commutant $\SU(6)$ of the gauged $\SU(2)$ inside $E_6$.

This observation suggests that a similar decomposition into sectors with an emergent weakly-coupled gauge multiplet can happen at a non-perturbative region of the Coulomb branch of an asymptotically-free $\cN=2$ gauge theory.
We study a simple class of such theories, namely $\SU(N)$ theory with an even number, $N_f=2n$, of  flavors with $N_f<2N$.
We will see that for $N_f\ge 4$ the theory at the maximally-singular point on the Coulomb branch has a dual description with an infrared-free $\SU(2)$ gauge multiplet coupled to two isolated superconformal theories.

Our interpretation of the physics at this point differs from \EguchiVU.
Furthermore, this reinterpretation allows us to reexamine the renormalization group flow between these theories \ShapereUN.  We conclude that in accordance with Cardy's conjectured $a$-theorem \CardyCWA\ (see also \refs{\BarnesJJ,\MyersXS})
the conformal central charge $a$ decreases along the renormalization group flow.

The rest of the paper is organized as follows.
Section 2 contains the main analysis of the $\SU(N)$ theory with $2n$ flavors.
In Sec.~2.1, the curve is rewritten to the form adapted to the study around the maximally-singular point.  In Sec.~2.2, we study the scaling procedure toward the maximally-singular point.  We will see that the curve develops a long tubular region connecting two almost decoupled sectors.
In Sec.~2.3, 2.4 and 2.5 the tubular region and the two sectors are analyzed in detail.  The analysis is then summarized in Sec.~2.6.
Section 3 consists of two checks and one application:
we study the beta function of the infrared-free magnetic $\SU(2)$ in Sec.~3.1
and the Higgs branch in Sec.~3.2.  The conformal central charge $a$ is studied in Sec.~3.3.  We conclude with a short discussion in Section 4.

\newsec{Analysis}
\subsec{Rewriting of the curve}
We start from the curve of $\SU(N)$ gauge theory with an even number $N_f=2n$ of flavors
for $N>n$ as determined in \refs{\ArgyresFW,\HananyNA}:
\eqn\curveA{
y^2=(\hat x^N+ \hat u_2 \hat x^{N-2} + \cdots +  \hat u_N)^2 - \Lambda^{2N-2n} \prod_{i=1}^{2n} (\hat x+ \hat m_i).
}   Here, $\hat u_k$'s are the coordinates of the Coulomb branch and $\hat m_i$'s are the masses of the hypermultiplets.   The central charge of BPS particles  is determined by the differential
\eqn\differentialA{
\lambda = \hat x \, d \log {y+P\over y-P}, \qquad \hbox{where} \qquad
P(\hat x)=\hat x^N+  \hat u_2  \hat x^{N-2} + \cdots +  \hat u_N
}

First we split the mass parameters $\hat m_i$ associated to the $\U(2n)$ symmetry into the $\U(1)$ part, $u_1$, and the $\SU(2n)$ part, $m_i$ (such that $\sum   m_i=0$) by shifting $\hat x$ and $\hat u_i$ in \curveA:
\eqn\curveB{
y^2=( x^N+   u_1   x^{N-1} +   u_2    x^{N-2} +   \cdots +    u_N)^2 - \Lambda^{2N-2n} \prod_{i=1}^{2n} (  x+  m_i)
}
Note that $u_1$ is a parameter and not a coordinate on the moduli space of vacua.  A further rewriting makes it to
\eqn\curveC{\eqalign{
y^2&=(   x^N+   u_1   x^{N-1} +   u_2    x^{N-2} +   \cdots +    u_N)^2 - \Lambda^{2N-2n}   x^{2n}
- \sum_{k=2}^{2n} c_k   x^{2n-k} \cr
&=(   x^N+ \cdots  +   u_{N-n}\new    x^{n} +   \cdots +    u_N)\times  \cr
&\qquad (   x^N+ \cdots  + ( 2\Lambda^{N-n} +   u_{N-n}\new )   x^{n} +   \cdots +    u_N) - \sum_{k=2}^{2n} c_k   x^{2n-k}
}}
Here, the parameters $c_k$ are the Casimir invariants of the $\SU(2n)$ mass, and $  u_{N-n}=\Lambda^{N-n} +   u_{N-n}\new$.
In the following we drop the superscript $\new$, hoping no confusion arises.

\subsec{Scaling limit}
When all $u_i$ and $c_k$ are small compared to $\Lambda$, the curve is approximated by
\eqn\limitcurve{
y^2 \approx (x^{N-n}+2\Lambda^{N-n}) x^{N+n}.
}
We call this point in the Coulomb branch of the $\SU(N)$ theory with $2n$ flavors {\it the EHIY point} \EguchiVU.

On the $x$ plane, there are $N-n$ branch points at $|x|\sim |\Lambda|$
and $N+n$ branch points at $|x|\sim 0$.
It is thus tempting to scale toward the small $x$ region and to set  the scaling dimensions of $x$ and $y$ to satisfy $[x]:[y]=(N+n):2$.
The differential in the small $x$ region can be approximated by
\eqn\limitdifferential{
\lambda \approx y dx / x^n .
}
Demanding $[\lambda]=1$ then determines $[x]$ and $[y]$, as was done in \EguchiVU.
However this assigns $[c_3]\ne 3$ when $n> 1$, or equivalently when $N_f> 2$.  This is in contradiction with the argument in \ArgyresXN\ that the scaling dimensions of mass parameters associated to non-Abelian flavor symmetry do not acquire anomalous dimensions.
Therefore, we need to re-analyze the situation when $n>1$.

Let us study more carefully the geometry of the small $x$ region of the curve,
where we have $N+n$ branch points.
We define $\tilde y=y/x^{n-1}$ and write the curve as
\eqn\curve{\eqalign{
\tilde y^2&= (x^{N-n+2}+ \cdots+ u_{N-n+1} x + u_{N-n+2}
+ {u_{N-n+3}\over x} + \cdots + {u_N\over x^{n-2}}  ) \cr
&\qquad \qquad \times (x^{N-n}+ \cdots+ (2\Lambda^{N-n} + u_{N-n} ) + {u_{N-n+1}\over x}+  { u_{N-n+2}\over x^2}
 + \cdots + {u_N\over x^{n}}  )  +\cr
&\qquad \qquad \qquad +c_2 + {c_3\over x}+\cdots+ {c_{2n}\over x^{2n-2}}.
}}
The differential is $\lambda\approx \tilde y dx/x$.

We anticipate a scaling limit of the curve \curve\ such that $u_j$ and $c_k$ are scaled to zero in a prescribed way, but different subsectors are obtained with different scalings of $x$.

We would like to keep $[c_k]=k$.  This motivates us to introduce $\epsilon_A\ll 1$ and scale $c_k \sim O(\epsilon_A^k)$.  This immediately implies that the scaled curve depends on $c_k$ only when $x \sim O(\epsilon_A)$, and therefore we take
\eqn\choiceA{
u_{N-n+2} \sim O(\epsilon_A^2), \quad
u_{N-n+3} \sim O(\epsilon_A^3), \quad
\ldots,\quad
u_{N} \sim O(\epsilon_A^n).
}
In order to account for another sector in which $x$ is scaled differently, let $x \sim O(\epsilon_B)$ and
\eqn\choiceB{
u_{1} \sim O(\epsilon_B), \quad
u_{2} \sim O(\epsilon_B^2), \quad
\ldots,\quad
u_{N-n+2} \sim O(\epsilon_B^{N-n+2}).
}
Clearly, consistency of \choiceB\ and \choiceA\ demands
\eqn\epsilonAB{\epsilon_A^2=\epsilon_B^{N-n+2}}
and hence $\epsilon_A \ll \epsilon_B \ll |\Lambda|\sim 1$.

The curve has three interesting regions:
\item{1.} For $|x|\sim \epsilon_A$ the curve is
\eqn\partA{\eqalign{
\tilde y^2&\approx( u_{N-n+2} + {u_{N-n+3}\over x} + \cdots + {u_N\over x^{n-2}}  ) \cr
&\qquad \times ( 2\Lambda^{N-n}+ {u_{N-n+2} \over x^2}+ {u_{N-n+3}\over x^3} + \cdots + {u_N\over x^{n}}  )
 +c_2 + {c_3\over x}+\cdots+ {c_{2n}\over x^{2n-2}},
}}
and it has $2n-2$ branch points.
\item{2.} For $|x|\sim \epsilon_B$ the curve is
\eqn\partB{
\tilde y^2\approx 2\Lambda^{N-n} (x^{N-n+2}+ \cdots+ u_{N-n+1} x + \check u_{N-n+2}),
}
where $\check u_{N-n+2}=u_{N-n+2} + c_2/(2\Lambda^{N-n})$.
It has $N-n+2$ branch points.
\item{3.} Between these two regions, $\epsilon_A\ll |x|\ll \epsilon_B$, the curve is trivial
\eqn\partC{ \tilde y^2\approx 2\Lambda^{N-n}  \check u_{N-n+2}.}
There are no branch points in this region.

\noindent
Let us discuss these three regions.

\subsec{$ \epsilon_A \ll |x| \ll \epsilon_B$}
This is a tubular region with the trivial curve \partC.
It gives a BPS vector multiplet with BPS central charge
\eqn\electric{
\oint_{|x|=\hbox{\sevenrm const.}} \lambda = 2\pi i a \qquad\hbox{where}\qquad
a^2= 2\Lambda^{N-n} \check u_{N-n+2}  \sim O(\epsilon_A^2).
}

There is also a BPS geodesic connecting a branch point at $|x|\sim \epsilon_A$
and another at $|x|\sim \epsilon_B$. This gives a massive hypermultiplet, whose BPS central charge is
\eqn\magnetic{
\int_{x\sim \epsilon_A} ^{x\sim \epsilon_B} \lambda =  a \times \left[{N-n \over N-n+2} (\log \epsilon_A)+ \hbox{const.} \right] .
}

When $\epsilon_A$ is very small, these represent the physics of a weakly-coupled $\SU(2)$ gauge multiplet broken by the adjoint vev $a$: the particle \electric\ is the W-boson, and the particle \magnetic\  is the 't Hooft-Polyakov monopole.
The ratio of the mass of the monopole to that of the W-boson grows as $\epsilon_A$ is lowered; therefore the $\SU(2)$ group is infrared free.
The coupling constant at the scale $a\sim \epsilon_A$ is
\eqn\running{
\tau = {1 \over 2\pi i} {N-n \over N-n+2} \log (\epsilon_A).
}
Equivalently, the one-loop running $b_0$ of the $\SU(2)$ gauge group is given by
\eqn\bee{
b_0={N-n \over N-n+2}
} which is positive.

\subsec{$|x|\sim \epsilon_B$}

The physics in this region is governed by the curve \partB. Note that the theory only depends on $\nu=N-n+1$. In fact, this is the maximally conformal point of $\SU(\nu)$ gauge theory with two hypermultiplets in the fundamental representation, which has $\U(2)$ flavor symmetry.
Let us call this theory $S_\nu$.
For $\nu=2$ this is the point studied in \ArgyresXN;
for $\nu>2$ this was studied in \EguchiVU.
The scaling dimensions are given by
$[u_k]={2k \over N-n+2}$.
$u_1$ is the $\U(1)$ mass parameter;
$\check u_{\nu+1}$ is the $\SU(2)$ mass parameter, which has the correct scaling dimension 2.

The flavor symmetry central charge $k$ of the $\SU(2)$ part\foot{As is customary, $k$ is normalized so that a hypermultiplet in the fundamental representation contributes by $2$ to $k$.
} can be calculated using the methods presented in  \ShapereZF, and is given by
\eqn\kOfAD{
k={4\nu \over \nu+1} = {4(N-n+1)\over N-n+2}.
}  For $\nu=2$ it was also determined in \AharonyDJ, see Table 2 therein.

\subsec{$|x|\sim \epsilon_A$}

The physics in this region is governed by the curve \partA, and depends only on $n$.
It can be found by taking a strong-coupling limit of $\SU(n)$ gauge theory with $N_f=2n$ flavors. The curve was determined in \refs{\ArgyresFW,\HananyNA} and is  given by
\eqn\conformalcurve{
y^2=(\hat x^{n}+ \hat u_2 \hat x^{n-2} + \cdots +  \hat u_{n})^2 - f(\tau) \prod_{i=1}^{2n} (\hat x+ \hat m_i).
}
Here $f(\tau)$ is a certain modular function given in the references just cited. Its only property we need is $f(\tau=1)=1$.

One can analyze the strong coupling limit  $\tau\to 1$  by repeating what we presented in Sec.~2.1 and 2.2 almost verbatim,
and we obtain the curves \partA\ and \partB\ with $N=n$.
In this particular case, the curve \partB\ produces just a hypermultiplet, and we conclude that the $\SU(n)$ theory with $2n$ flavors at the strongly-coupled limit $\tau\to 1$ has an S-dual description, consisting of
\item{$\bullet$} an $\SU(2)$ gauge multiplet, coupled to
\item{$\bullet$} a hypermultiplet in the doublet, coming from the curve \partB,
which has $\SU(2)\times \U(1)$ flavor symmetry, and
\item{$\bullet$}  a strongly-coupled theory coming from the curve \partA,
which has $\SU(2)\times \SU(2n)$ flavor symmetry. Let us call this theory $R_n$.

\medskip

For $n=2$, this S-duality  is the standard one of  $\SU(2)$ theory with four flavors.
Therefore $R_2$  is a free theory of $N_f=3$ doublets of $\SU(2)$. This has $\SU(2)\times \SO(2N_f)\simeq \SU(2)\times \SU(4)$ flavor symmetry.

For $n=3$, this S-duality is the one first found in \ArgyresCN. Therefore $R_3$ is the isolated rank-1 theory with the flavor symmetry $E_6 \supset \SU(2)\times \SU(6)$, originally found in \MinahanFG.

For $n\ge 4$, this S-duality is of the type whose analysis was initiated by \GaiottoWE; this particular case was studied in detail by \ChacaltanaKS.
$R_n$ is an isolated superconformal field theory with the flavor symmetry $\SU(2)\times \SU(2n)$. The flavor symmetry central charge $k$ of $\SU(2)$ part is $k=6$.

\subsec{Summary}
Combining the analyses in Sec.~2.3, 2.4 and 2.5, we conclude that
$\SU(N)$ gauge theory with $N_f=2n$ flavors at the EHIY  point is described by
\item{$\bullet$} a number of decoupled $\U(1)$ vector multiplets,
\item{$\bullet$} an infrared-free $\SU(2)$ gauge multiplet, coupled to
\item{$\bullet$} a strongly-coupled superconformal theory $S_{N-n+1}$, {\it i.e.} the maximally conformal point of $\SU(N-n+1)$ theory with two flavors, which has $\U(1)\times \SU(2)$ flavor symmetry, and
\item{$\bullet$} another strongly-coupled superconformal theory $R_n$, {\it i.e.} a component of the S-dual of $\SU(n)$ theory with $2n$ flavors, which has $\SU(2)\times \SU(2n)$  flavor symmetry.

\medskip

The main point (which differs from \EguchiVU) is that we find, when $N_f=2n \ge 4$, two almost-decoupled sectors $S_{N-n+1}$ and $R_n$ coupled by an infrared-free $\SU(2)$.
In the curve \partA, considered in itself, the parameter $\check u_{N-n+2}$ represents the squared  mass parameter of the $\SU(2)$ flavor symmetry of the theory $S_{N-n+1}$.
Similarly,  in the curve \curveB, $\check u_{N-n+2}$ is the squared mass parameter of the $\SU(2)$ flavor symmetry of the theory $R_n$.
The dynamical $\SU(2)$ vector multiplet coming from the tubular region \partC\ couples to the diagonal combination of these $\SU(2)$ flavor symmetries, and indeed the mass of the vector boson is given by the square root of $\check u_{N-n+2}$, as shown in \electric.

\newsec{Two checks and one application}
Here we perform two easy checks and an application of the analysis presented in the previous section.

\subsec{One-loop beta function}
The first check is the one-loop beta function of the $\SU(2)$ gauge group.
When an $\SU(N)$ gauge multiplet is coupled to a system with $\SU(N)$ flavor symmetry with central charge $k$, the one-loop coefficient $b_0$ is given by
\eqn\generalbzero{
b_0={k\over 2}-2N
} where the second term is the contribution from the $\cN=2$ vector multiplet.
The theory $S_{N-n+1}$ has $k$ given by \kOfAD, while the theory $R_n$ has $k=6$. We correctly reproduce $b_0$ given in \bee, which was calculated directly from the  curve of the whole system.

\subsec{Higgs branch}
The second check is the Higgs branch of the system.  As shown in \ArgyresEH, the EHIY point on the Coulomb branch touches the origin of the Higgs branch of quaternionic dimension $n^2$ of the form
\eqn\totalHiggs{
\H^{2n\times n} \hkq  \U(n),
} where the symbol $M \hkq G$ stands for the hyperk\"ahler quotient, or equivalently the Higgs branch of an $\cN=2$ theory with gauge group $G$ coupled to hypermultiplets parameterizing $M$.

In the description given in Sec.~2.6, the Higgs branch has the form
\eqn\HiggsInThisDescription{
( \Higgs(S_{N-n+1}) \times \Higgs(R_n)  ) \hkq \SU(2),
} where $\Higgs(X)$ stands for the Higgs branch of theory $X$.
Let us check that this  agrees with \totalHiggs.
Firstly, from \ArgyresEH, we know
\eqn\SHiggs{
\Higgs(S_\nu) = \H^2 \hkq \U(1) = \C^2/\Z_2.
}
Secondly, from the S-duality of $\SU(n)$ theory with $2n$ flavors, we know\foot{This was explicitly checked for $n=3$ in \GaiottoNZ.}
\eqn\RHiggs{
(\H^2 \times \Higgs(R_n) ) \hkq \SU(2)
= ( \H^{2n\times n}) \hkq \SU(n).
}
Therefore, we have
\eqn\foo{\eqalign{
( \Higgs(S_{N-n+1}) \times \Higgs(R_n)  ) \hkq \SU(2)
&= ( (\H^2\hkq \U(1)) \times \Higgs(R_n)  ) \hkq \SU(2)\cr
&= ( \H^2  \times \Higgs(R_n)  ) \hkq \SU(2) \hkq \U(1)\cr
&=  ( \H^{2n\times n}) \hkq \SU(n) \hkq \U(1)\cr
&=  ( \H^{2n\times n}) \hkq \U(n) ,
}} thus reproducing \totalHiggs.

\subsec{Renormalization-group flow and the central charge $a$}

Following \ShapereUN, we would like to examine the renormalization group flow between the EHIY points of $\SU(N+1)$ theory with $N_f=2n$ flavors and that of $\SU(N)$ theory with the same number of flavors.  In particular, we are interested in the conformal central charge $a$.  The reinterpretation of the EHIY point leads us to recalculate $a$.

Clearly, the renormalization group flow is essentially between $S_{\nu+1}$ and $S_{\nu}$, while
the $R_n$ theory is a spectator. The central charge $a$ of the theory $S_{\nu}$ was calculated in \ShapereUN:
\eqn\Aeven{
a=\cases{\displaystyle
{\nu\over 4}-{1\over 16}+{1\over 16(\nu+1)}
&  for even $\nu$,  \cr
\displaystyle
{\nu\over 4}-{1\over 6}& for odd $\nu$.}}
As the low energy limit of $\SU(\nu)$ theory with two flavors,
we have additional $\nu/2-1$ free $\U(1)$ vector multiplets
when $\nu$ is even, and $(\nu-1)/2$  when $\nu$ is odd. One free $\U(1)$ vector multiplet contributes $5/24$ to $a$, and therefore the total $a$ is given by
\eqn\AevenX{
a=\cases{\displaystyle
{17 \nu\over 48}-{13\over 48}+{1\over 16(\nu+1)}
 & for even  $\nu$, \cr
\displaystyle {17\nu\over 48}-{13\over 48}
& for odd $\nu$.}}
The conformal central charge  $a$ decreases as $\nu$ decreases, in accordance with \CardyCWA.

\newsec{Discussions and conclusions}
We are often interested in identifying the low energy physics at the vicinity of a point $\cP$ on the Coulomb branch.
For generic $\cP$ the low energy theory consists of a number of $\U(1)$ vectors with massive charged hypermultiplets, and the analysis is trivial.
At some special points, we find massless charged hypermultiplets coupled to infrared-free gauge multiplets.
Here we see monodromies around $\cP$ and possibly a Higgs branch emanating from the point.
This case is still relatively easy to understand and by now it is standard.
More interesting is the situation in which the low-energy limit at $\cP$ is an interacting theory.
In this case we should find an appropriate scaling toward $\cP$ and identify the interacting scaling theory.

Finding such a scaling is not always obvious, and we do not yet have a clear and straightforward procedure to find the correct scaling at a given point $\cP$.
Instead, we need to come up with an informed guess, which we subject to various consistency conditions.

One constraint is that all the non-Abelian flavor currents, and therefore the corresponding mass parameters, should have their canonical dimensions.
This requirement can lead to a scaling limit which splits the curve, and therefore the physics, into sectors which are weakly coupled by a gauge multiplet.  Then the next step is to identify each sector as one of the already known theories.  Once this step is done, we can perform further consistency checks, by comparing the description as the almost decoupled system against the original theory.  For example, the behavior of the gauge coupling can be studied in two ways, first by calculating the one-loop contributions from the decoupled sectors, and second by studying the BPS central charges from the full curve of the original system.  These two approaches should give the same beta function.  Another check is associated with the Higgs branch.  It can be determined both from the UV description and from the IR description.
These two calculations should again agree, because the Higgs branch is known not to be modified by the gauge dynamics \ArgyresEH.

It is, however, important to keep in mind that these are necessary conditions, but they might not be sufficient.  There is no guarantee that these considerations always lead to a unique answer.

In this paper we studied the EHIY point on the Coulomb branch of $\SU(N)$ theory with $2n$ flavors.  The analysis of this point clearly demonstrates that the limit can be subtle.  Motivated by the criteria above we proposed the scaling \choiceA--\epsilonAB.  The main subtlety in our answer is that the limit is such that the curve splits into sectors.
We found that when $n>1$ the infrared theory consists of an infrared-free $\SU(2)$ theory coupled to two isolated superconformal theories:
one is the theory $S_{N-n+1}$ at the EHIY point of the $\SU(N-n+1)$ theory with two flavors, the other is the theory $R_n$, which is a component of the S-dual of the $\SU(n)$ theory with $2n$ flavors in the limit $\tau\to 1$.

A possible tool in analyzing this problem is the construction of the $\cN=2$ theory as the six-dimensional $\cN=(2,0)$ theory compactified on a Riemann surface with punctures.
Perhaps this viewpoint can help determine the correct scaling limit.
In many examples, the scaling procedure applied to the punctured Riemann surface gives a direct six-dimensional construction of the corresponding isolated superconformal theories, as in \GaiottoHG.
We hope to come back to these problems in the future.

\bigskip \goodbreak

\centerline{\bf Acknowledgements}
NS is supported in part by DOE grant DE-FG02-90ER40542.
DG and YT are supported in part by NSF grant PHY-0969448.
YT is additionally supported by the Marvin L. Goldberger membership at the Institute for Advanced Study, and by World Premier International Research Center Initiative (WPI Initiative),  MEXT, Japan through the Institute for the Physics and Mathematics of the Universe, the University of Tokyo.

\listrefs
\bye